\documentclass[twocolumn,showpacs,prc,superscriptaddress]{revtex4-1}
\usepackage{graphicx}% Include figure files
\usepackage{dcolumn}% Align table columns on decimal point
\usepackage{bm}% bold math

\begin{document}

\title{No influence of a N=126 Neutron Shell Closure in Fission Fragment Mass Distributions \\}

\author{A. Chaudhuri} 
\author {T. K. Ghosh} \email{ E-mail: tilak@vecc.gov.in}
\author{K. Banerjee}
\author{S. Bhattacharya} \thanks{Raja Ramanna Fellow}
\author{Jhilam Sadhukhan}
\author {S. Kundu}
\author{C. Bhattacharya}
\author {J. K. Meena}
\author {G. Mukherjee}
\author {A. K. Saha}
\author{Md. A. Asgar}
\author{A. Dey}
\author {S. Manna}
\author {R. Pandey}
\author {T. K. Rana}
\author {P. Roy}
\author {T. Roy}
\author {V. Srivastava} 
\address{Variable Energy Cyclotron Centre,  1/AF, Bidhan  Nagar,
  Kolkata  700064, India}

	\author{P. Bhattacharya}  
	\address{Saha Institute of Nuclear Physics,  1/AF, Bidhan  Nagar,
  Kolkata 700064, India.}
\author {D. C. Biswas}
\address {Nuclear Physics Division, Bhabha Atomic Research Centre, Mumbai 400085, India}

\author {B. N. Joshi}
\address {Nuclear Physics Division, Bhabha Atomic Research Centre, Mumbai 400085, India}
\author {K. Mahata}
\address {Nuclear Physics Division, Bhabha Atomic Research Centre, Mumbai 400085, India}
\author {A. Shrivastava}
\address {Nuclear Physics Division, Bhabha Atomic Research Centre, Mumbai 400085, India}
\author {R. P. Vind}
\address {Nuclear Physics Division, Bhabha Atomic Research Centre, Mumbai 400085, India}
\author {S. Pal}
\address {Tata Institute of Fundamental Research, Mumbai 400005, India}
\author {B. R. Behera}
\address {Department of Physics, Panjab University, Chandigarh 160014, India}
\author {Varinderjit Singh}
\address {Department of Physics, Panjab University, Chandigarh 160014, India}
\date{\today}


\begin{abstract}

Mass distributions of the fragments in the fission of $^{206}$Po and the N=126 neutron shell closed nucleus $^{210}$Po have been measured. No significant deviation of mass distributions has been found between $^{206}$Po and $^{210}$Po, indicating the absence of shell correction at the saddle point in both the nuclei, contrary to the reported angular anisotropy and pre-scission neutron multiplicity results. This new result provides benchmark data to test the new fission dynamical models to study the effect of shell correction on the potential energy surface at saddle point.

\end{abstract}

\maketitle

The role of nuclear shell effects on various nuclear reaction processes as a function of excitation energy, specially for the nuclei  around the shell closure, has currently remained an issue of intense discussions \cite{RoutPRL,aradhana,schmitt,golda,Sagaidak09,karpov,mahata15,Abhirup,Prasad}. Apart from the basic understanding point of view, a large part of the recent activities were concentrated on the study of shell effect in fission of heavy nuclei with the aim to unveil the relationship between nuclear structure and nuclear stability. 

There have been immense efforts, both theoretical and experimental, to address the burning question whether nuclear shell effects survives around the saddle point. The theoretical efforts are concentrated on calculating the potential energy surfaces (PES) in a multi-dimensional space. It is found, in general, while the ground state mass is strongly influenced by the shell correction, the saddle point mass should be rather close to its macroscopic value \cite{karpov,myers}. In contrast, a few recent experimental studies indicated rather strong effect of shell correction at the saddle point, particularly around N=126 shell closed nuclei \cite{mahata06,aradhana,golda}. An anomalous increase in fission fragment angular anisotropy was observed  in the fission of  $^{210}$Po at excitation energy $ \sim $40-60 MeV, which was conjectured as an indirect evidence of shell correction at saddle due to neutron shell closure at N=126 \cite{aradhana,mahata06}. The pre-scission neutron multiplicity data  for $^{206, 210}$Po also indicated the requirement of substantial shell correction not only in  shell closed $^{210}$Po but also in $^{206}$Po \cite{golda}. These results, which are apparently indicative of a much stronger role of nuclear structure in fission process is bound to have implications on all future studies of the fission, and  particularly will have vital impact on the production of spherical super heavy nuclei around the next closed neutron shell at N = 184. Therefore, it warrants independent attempt to estimate the role of shell correction at saddle point in the same mass/excitation energy region where the deviations were observed. 

The analysis of both angular anisotropy and pre-scission neutron data mentioned above \cite{aradhana,mahata06,golda}, were carried out within the framework of the well established statistical models \cite{Vandenbosch, Gavron} which are fairly successful in explaining the gross features of the binary fission of a statistically equilibrated compound nucleus. As the gross effects of shell structure are already taken care of in the model calculations through the shell corrected level density term, any departure of the measured evaporation residue yield, anisotropy or pre-scission neutron multiplicity from the corresponding model predicted values may be construed either as the manifestation of shell structure on the PES or as the contributions from other non-compound fission channels. However, the robustness of the statistical model predictions was recently called into question \cite {schmitt}. With the advent of dynamical calculations using stochastic Langevin equation, Schmitt {\sl et al.} \cite {schmitt} showed that the angular anisotropy and neutron data \cite{aradhana,golda}, mentioned above, could well be explained with purely macroscopic potential energy landscape without considering any shell correction at saddle point. The prevailing dramatic ambiguity thus necessitates an immediate evaluation of the problem through a new experimental observable, the fission fragment mass distribution, which would probe the PES directly at the saddle point, as the mass ratio of the emitted fragments largely depends on the structure of the potential energy surface at the saddle point \cite{Itkis91}. 

In this Communication, we report a measurement of fission fragment mass distributions for the fissioning nuclei $^{206, 210}$Po to look for signatures (if any) of shell correction on the potential energy surface at the saddle. No significant deviation of mass distribution was found between $^{206}$Po and $^{210}$Po and both the distributions could be explained using realistic macroscopic potential only, contrary to the reported angular anisotropy and pre-scission neutron multiplicity results. 
 
\begin{figure}
\includegraphics*[scale=0.30, angle=0]{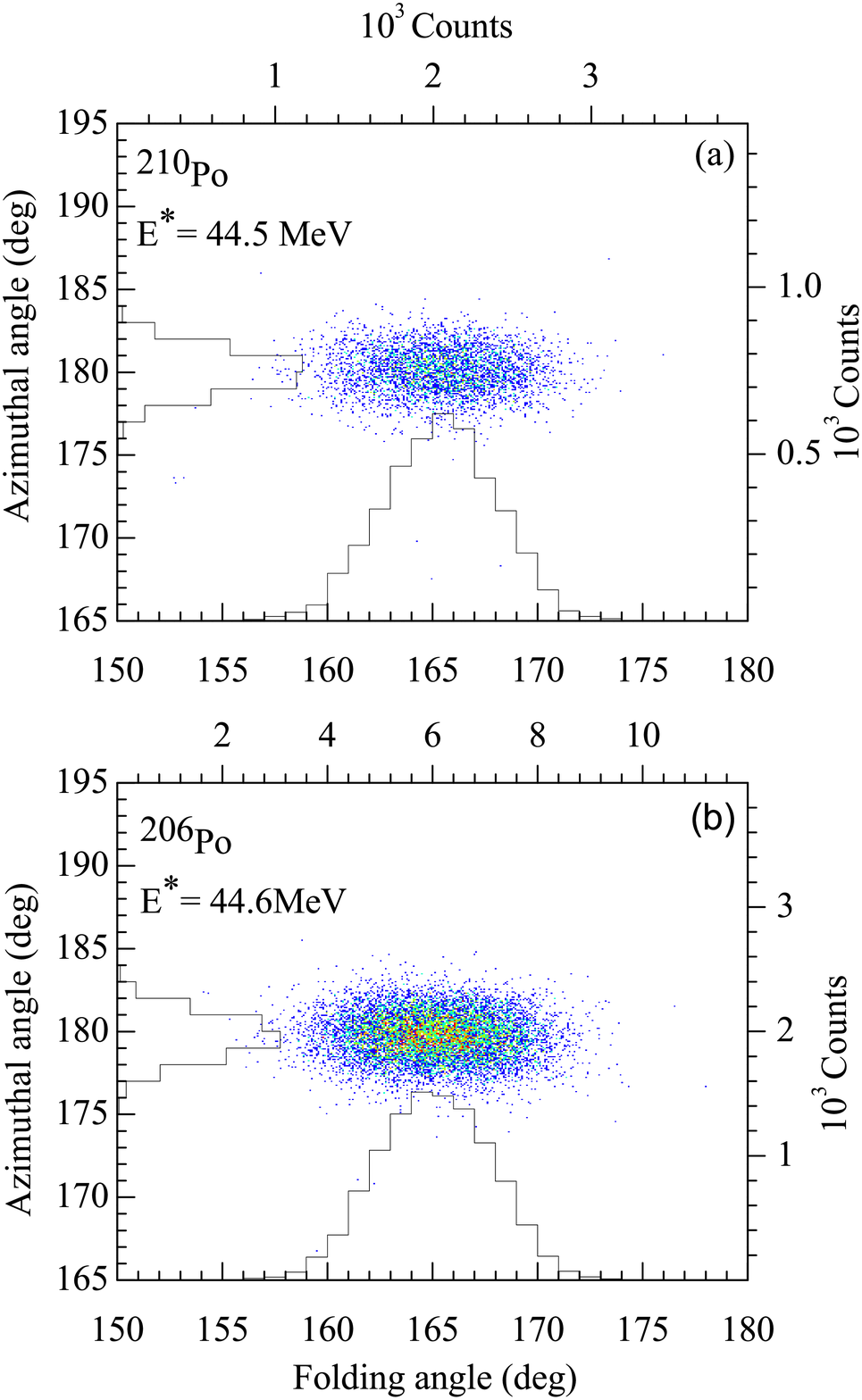}
\caption{\label{fig:fig1}~(Color online) Distributions of folding angles of complimentary fission fragments for the reactions (a) $^{12}$C+$^{198}$Pt and (b) $^{12}$C+$^{194}$Pt at similar excitation energies.} 
\end{figure}

The experiment was performed at the BARC-TIFR Pelletron facility at Mumbai, India with bunched beam of $^{12}$C (58 - 78 MeV) on (96.5\% enriched) isotopes of $^{194}$Pt of thickness 260 $\mu$g/$cm^2$ (carbon backing 20 $\mu$g/$cm^2$)  and $^{198}$Pt (91.6\% enriched) of thickness 170 $\mu$g/$cm^2$ (10 $\mu$g/$cm^2$). Targets were mounted at an angle of 45$^{\circ}$ to the beam. Fission fragments were detected with two large area position sensitive MWPC \cite{myNIM}. The detectors were placed at 48 cm and 37 cm from the target on either side of the beam axis. The centre of the forward detector was kept at an angle of 45$^{\circ}$ and the backward detector at 121$^{\circ}$ to the beam. The operating pressures of the detectors were maintained at 3 torr of iso-butane gas.  At this low pressure, the detectors were almost transparent to elastic and quasi-elastic particles. We measured the flight times of the fragments, the coordinates of the impact points of the fragments on the detectors ($\theta,\phi$), and the energy losses in the gas detectors. From these measurements, we extracted the masses of the correlated fission events and the transferred momentum to the fissioning system. Beam flux monitoring as well as normalization were performed using the elastic events collected by a silicon surface barrier detector placed at 15$^{\circ}$ to the beam and Faraday cup.

Typical folding angle distributions of all fission fragments (FF) in the two reactions measured at near Coulomb barrier energies are shown in Fig. 1. It is found that the peak of the folding angle distribution in each of the reactions is consistent with the expected value for complete transfer of momentum of the projectile. This, along with the symmetric shape of the distribution clearly show that there is no admixture of transfer induced fission fragments and all the fragments are originated in the fusion-fission reactions. 

\begin{figure}
\includegraphics*[scale=0.3, angle=0]{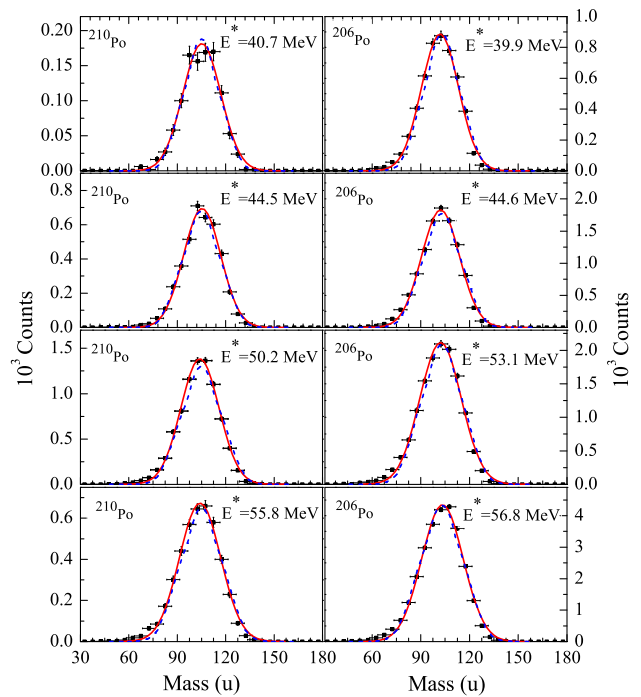}
\caption{\label{fig:fig2}~(Color online) Measured mass distributions of fission fragments at different excitation energies. Fittings using single Gaussian are shown by solid (red) lines. Theoretical calculations (refer to text) are shown by the dashed (blue) lines.} 
\end{figure}

The fission fragments are well separated from elastic and quasi-elastic reaction channels, both from the time correlation and energy loss spectra in the detectors. The masses were determined from the difference of the time of flights, polar and azimuthal angles, momenta, and the recoil velocities for each event. The mass distributions of fission fragments were determined following a procedure described in details earlier \cite{myNIM,mythesis}. The mass resolution achieved $\sim$ 5 u. Since the targets were not 100\% pure and as there is no way to distinguish the origin of the fission fragments detected by the detectors (whether they originated from the compound nuclei formed in fusion of the projectile and the main targets $^{194,198}$Pt or their isotopic impurities), we estimated the effect of the impurities by assuming proportionate number of the actual events coming from the isotopic impurities chosen randomly through a time seeded uniform random number generator. To estimate the dispersion in the variance of mass distribution due to the presence of isotopic impurities, the above process was repeated 2000 times. It is found that the dispersion in the variance of mass distribution due to impurities were negligibly small.

\begin{figure}
\includegraphics*[scale=0.4, angle=0]{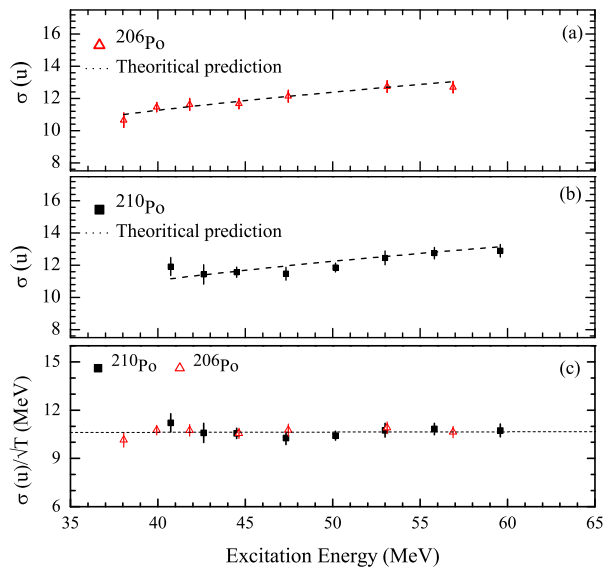}
\caption{\label{fig:fig3}~(Color online) (a),(b):Variation of the standard deviation of the fitted symmetric mass distribution with excitation energy. The calculated standard deviations are shown by dotted lines. (c) Variation of the standard deviation normalized by the saddle temperature. The dashed line (constant value) is a guide to the eye.} 
\end{figure}

Typical mass distributions of the fission fragments, measured at similar excitation energies for the two reactions are shown in Fig. 2.  The solid (red) line is a single Gaussian fit to the data. The good fits to the experimental data using a single Gaussian function in both the reactions are clearly confirming  that the FF mass distributions are completely symmetric having nearly identical shapes at all excitation energies. 

To have a further insight into the result, the standard deviations ($\sigma$) of the fitted mass distributions are plotted as a function of excitation energy in Figs. 3 (a),(b) for both the reaction channels of $^{12}$C+$^{194,198}$Pt. In the case of statistical fission of the compound nucleus, the standard deviation of the fragment mass distribution is known to follow the relation $\sigma = \sqrt{\frac{T}{k}}$, where $T$ is the temperature at the saddle point and $k$ is the stiffness parameter for the mass asymmetry degree of freedom \cite{Back96}. As shown in Fig. 3(c), the constant value of $\frac{\sigma}{\sqrt{T}}$ with excitation energy indicates purely statistical compound nuclear fission process in both the cases. The value of $k$ was found to be consistent with the comprehensive compilation of the data presented in reference \cite{Itkis}. The non-compound fission processes and/or the presence of shell correction, both of which would have triggered an anomalous variation of $\sigma$ \cite{BackPRC99,myPLB}, are therefore quite unlikely in either of the two reaction channels, which is clearly at variance with the earlier results for the same systems obtained using different probes \cite{aradhana,golda}.

\begin{figure}
\includegraphics*[scale=0.38, angle=0]{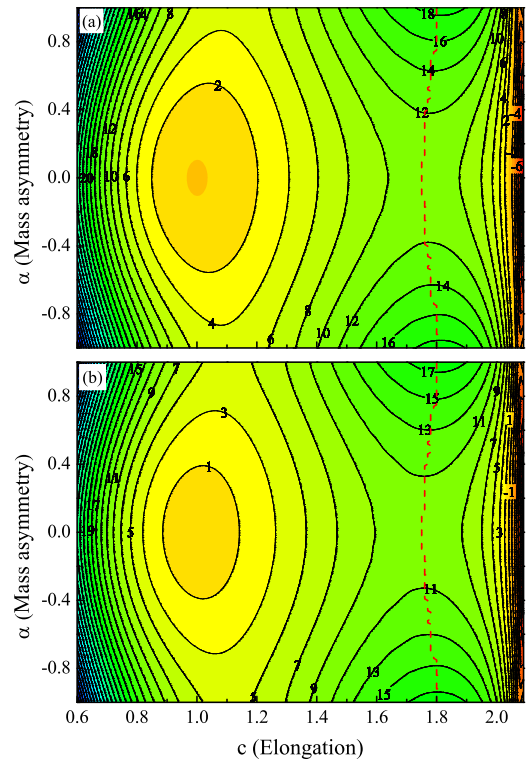}
\caption{\label{fig:fig4}~(Color online) Potential energy for the compound nuclei (a) $^{210}$Po and (b) $^{206}$Po relative to the ground state energy of FRLDM. All the contours are plotted at steps of 2 MeV. The (red) dashed line represents the computed saddle ridge.} 
\end{figure}

That the present non-observation of any appreciable anomaly in mass distribution (and vis-a-vis signature of shell correction) in either of the two systems $^{206,210}$Po is also justified theoretically, will be apparent from the following. According to the detailed theoretical calculation of PES \cite{moller}, fission barriers are single-peaked for the systems under consideration. So, an attempt was made to reproduce the measured mass distributions theoretically, considering only realistic macroscopic potential without any microscopic shell correction. We  calculated the PES using the Finite Range Liquid Drop Model (FRLDM) formula \cite{sierk, nix_sierk1, nix_sierk2}. The nuclear shapes were defined in two dimensions with “Funny Hill” \cite{funny_hill} parameters, elongation ($c$) and mass-asymmetry ($\alpha$). The fragment masses (M), corresponding to a particular combination of $c$ and $\alpha$, were decided by dividing the compound nucleus at the neck of the deformed shape. The calculated potential energy surfaces for $^{206}$Po and $^{210}$Po are plotted in Fig. 4. For each system, the saddle ridge, which defines the fission barrier $V(\alpha)$ as a function of $\alpha$, is shown by (red) dashed line. An estimate of the fission fragment mass distribution can be obtained from multi-dimensional Kramers’ formula for the fission width \cite{shang, tillack},

\begin{equation}
\Gamma_{f}=N(\alpha)exp(-V(\alpha)/T), 
\end{equation}
where, the coefficient $N(\alpha)$ depends on the detail structure of the potential profile and $T$ is the compound nuclear temperature calculated at the saddle point deformation. For the present calculation, $N(\alpha)$ was assumed to be independent of $\alpha$ and  we used a simplistic prescription of multiplying $V(\alpha)/T$ by a factor $B$ to take care of the dynamical effects \cite{jhilam}. It was found that constant values of $B$ (1.93 and 1.82) reproduced the experimental data very well for $^{206}$Po and $^{210}$Po, respectively (shown by dashed blue line in Fig. 2). The standard deviations of the theoretical mass distributions are also found to reproduce the experimental data reasonably well as shown (by dotted line) in Fig. 3.

The change in shape or width of the fission fragment mass distribution is a signal for the presence of shell correction at saddle point \cite{Itkis91}. It is clearly evident from our data that, so far as the fission fragment mass distributions are concerned, there is no anomaly between the two systems, $^{206}$Po and $^{210}$Po. Experimentally, both of them exhibit symmetric Gaussian-like mass distribution without any appreciable change of shape (width) over the whole range of excitation energy under consideration. Theoretical mass distributions, obtained using the PES without incorporating shell correction, were found to properly reproduce the respective experimental data in both cases.Thus, it is clear that the N=126 shell closure in $^{210}$Po does not affect the FF mass distribution.

\begin{figure}
\includegraphics*[scale=0.27, angle=0]{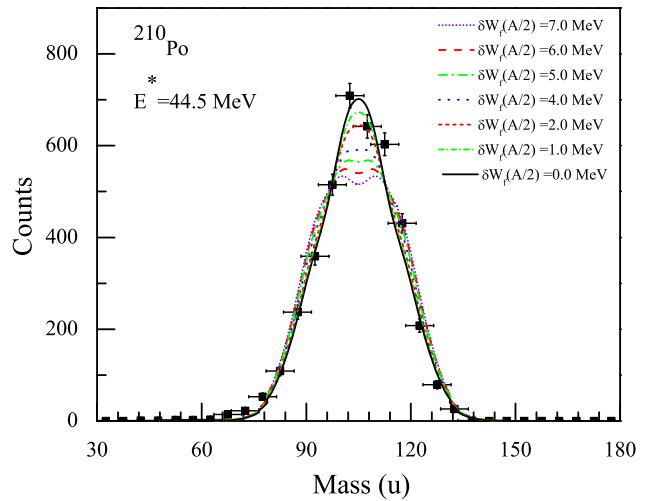}
\caption{\label{fig:fig5}~(Color online) Effect of shell correction in the saddle ridge on the mass distribution of fission fragments. $\delta W_{f}(A/2)$ is the shell correction at symmetry.} 
\end{figure}

In order to quantitatively workout the sensitivity of the fission fragment mass distributions on the magnitude of shell correction at saddle, the mass asymmetry dependent fission barrier was modified by adding a shell correction term $\delta W_{f}(M)exp[\lambda E^{*}]$, where $\lambda$ is the shell damping factor which was taken to be 0.054, and $\delta W_{f}(M)$  was taken in the empirical form \cite{Itkis91}
\begin{equation}
\delta W_{f}(M)= \delta W_{f}(A/2)exp[-\gamma(M-A/2)^{2}]. 
\end{equation}
The variation in mass distribution shape for different values of $\delta W_{f}(A/2)$ is displayed in Fig. 5. for a typical case of $^{210}$Po fission, where $\delta W_{f}(A/2)$ is the shell correction at symmetry, was varied over the range of 1-7 MeV. It is clear that the measured mass distribution can be best fitted with no shell corrections at saddle. It can be seen that at 7 MeV of shell correction at saddle, which was required to fit the neutron multiplicity data for the same system \cite{golda}, the theoretical mass distribution is clearly asymmetric as opposed to our experimental observation. It is worth mentioning here that even a variation in $\delta W_{f}(A/2)$ of 1 MeV produces perceivable change in the shape of the mass distribution.

The fission fragment mass distribution of $ ^{210}$Po, populated through $ ^{4}$He + $ ^{206}$Pb, was also reported by Itkis {\sl et al.} \cite{Itkis, Mulgin}. A comparative study of the width of the mass distribution at the only overlapping excitation energy showed slight suppression for the $ ^{4}$He + $ ^{206}$Pb reaction compared to our measurement owing to lower angular momentum carried by $ ^{4}$He as compared to $ ^{12}$C. A very weak structure in the mass distribution was reported by Itkis {\sl et al.}, which was however not observed in the present case. This may be due to the inherent sensitivity of our spectrometer. Regarding the contributions of other non-compound fission channels, it may be pointed out that their presence too would have appreciably broadened the mass distributions \cite{myPLB,kaushik}. The width of the distribution remained constant over the whole range of excitation energy (excluding temperature effect) is a clear indication that the contributions of non-compound channels are also minimal in the present case.

It may be mentioned that the fission fragment angular anisotropy was measured \cite{kripa} for the nucleus $^{213}$Fr which is also a  N=126  neutron-shell closed nucleus. Interestingly, any such appreciable deviation from statistical model predictions at similar excitation energies were not observed; and reanalysis of the $ ^{210}$Po data \cite{aradhana} including multi-chance nature of fission reduced the discrepancy in angular anisotropy. Systematic studies of angular anisotropies for different isotopes were carried out for a few other systems \cite{vigdor}; in none of the cases, large deviation of anisotropy were observed.  A recent calculation \cite{mahata15} for $ ^{210}$Po, however, advocated for minimal shell correction at saddle but required substantial dynamical effect to explain simultaneously the fission excitation function and neutron emission data. This reinforces our conviction that the anomalies in angular anisotropy and neutron multiplicity observed in the systems $^{206,210}$Po may not be attributed to the effect of neutron shell closure or shell correction at saddle point - it could be due to the inherent limitations of the implementations of statistical models \cite{schmitt}. 

In conclusion, the direct probe of fission fragment mass distribution does not show any signature of the modification of the potential energy surface at the saddle point due to the effect of N=126 neutron shell closure in $^{210}$Po. These results provide a benchmark for different models that are used to predict the fission barriers for the production of spherical super heavy nuclei around the next closed neutron shell at N = 184. The present findings merits further investigation for the other regions of neutron or proton shell closure. 

We are thankful to the staff members of the BARC-TIFR Pelletron for providing good quality of pulsed beam required for the experiment. Thanks are due to Dr. Santanu Pal for illumination discussions and Dr. Mishreyee Bhattacharya for critical reading of the manuscript. A.C., A.A., T.R. and V.S. acknowledge with thanks the financial support received as Research Fellow from the Department of Atomic Energy, Government of India. A.D acknowledges with thanks the financial support provided by the Science and Engineering Research Board, Department of Science and Technology, Government of India; and S. B. acknowledges with thanks the financial support received as Raja Ramanna Fellow from the Department of Atomic Energy, Government of India.

\end{document}